\documentclass[]{spie}  

\usepackage{amsmath,amsfonts,amssymb}
\usepackage{graphicx}
\usepackage[colorlinks=true, allcolors=blue]{hyperref}
\begin{document} 

\title{Imaging low-mass planets within the habitable zones of nearby stars with ground-based mid-infrared imaging}

\author[a,b,c]{Kevin Wagner} 
\affil[a]{Steward Observatory, University of Arizona, USA}
\affil[b]{NASA Hubble Fellowship Program $-$ Sagan Fellow}
\affil[c]{NASA Nexus for Exoplanet System Science $-$ Alien Earths Team}
\author[a,d]{Steve Ertel}
\affil[d]{Large Binocular Telescope Observatory, Tucson, AZ, USA}
\author[e]{Jordan Stone}
\affil[e]{Naval Research Laboratory, Washington, D.C., USA}
\author[a]{Jarron Leisenring}
\author[a,c,f]{D\'aniel Apai}
\affil[f]{Lunar and Planetary Laboratory, University of Arizona, USA}
\author[g]{Markus Kasper}
\affil[g]{European Southern Observatory, Garching, Germany}

\author[h]{Olivier Absil}
\affil[h]{Universit\'e of Li\`ege, Belgium}
\author[a,c]{Laird Close}
\author[h]{Denis Defr\`ere}
\author[a,i]{Olivier Guyon}
\affil[i]{Subaru Telescope, Hilo, Hawaii, USA}
\author[a,c]{Jared Males}

\authorinfo{Correspondence to kevinwagner@email.arizona.edu}

\pagestyle{plain}
\setcounter{page}{1} 
 
\maketitle

\begin{abstract}
Giant exoplanets on 10-100 au orbits have been directly imaged around young stars. The peak of the thermal emission from these warm young planets is in the near-infrared ($\sim$1-5 µm), whereas mature, temperate exoplanets (i.e., those within their stars' habitable zones) radiate primarily in the mid-infrared (mid-IR: $\sim$10 $\mu$m). If the background noise in the mid-IR can be mitigated, then exoplanets with low masses–including rocky exoplanets–can potentially be imaged in very deep exposures. Here, we review the recent results of the Breakthrough Watch/New Earths in the Alpha Centauri Region (NEAR) program on the Very Large Telescope (VLT) in Chile. NEAR pioneered a ground-based mid-IR observing approach designed to push the capabilities for exoplanet imaging with a specific focus on the closest stellar system, $\alpha$ Centauri. NEAR combined several new optical technologies–including a mid-IR optimized coronagraph, adaptive optics system, and rapid chopping strategy to mitigate noise from the central star and thermal background within the habitable zone. We focus on the lessons of the VLT/NEAR campaign to improve future instrumentation$-$specifically on strategies to improve noise mitigation through chopping. We also present the design and commissioning of the Large Binocular Telescope's Exploratory Survey for Super-Earths Orbiting Nearby Stars (LESSONS), an experiment in the Northern hemisphere that is building on what was learned from NEAR to further push the sensitivity of mid-IR imaging. Finally, we briefly discuss some of the possibilities that mid-IR imaging will enable for exoplanet science.


\end{abstract}


\section{INTRODUCTION}
\label{sec:intro}  

Earth-sized exoplanets have been indirectly observed through transit and radial velocity surveys [e.g., \citenum{Charbonneau2009,Jenkins2015,Berta2015,Dittmann2015,Anglada2016,Gillon2017}], while direct imaging has enabled the discovery of young super-Jovian planets on wide orbits [e.g., \citenum{Marois2010,Lagrange2010,Rameau2013,Macintosh2015,Chauvin2017,Keppler2018,Bohn2020}]. For the nearest stars, the habitable zones overlap with background-limited (and not contrast-limited) sensitivity regions of large telescopes operating in the mid-infrared (mid-IR). Such wavelengths also coincide with the thermal emission peak of temperate ($T\sim$ 300K) exoplanets. Therefore, low-mass exoplanets can be directly imaged with sufficient exposure time. In a three-week observing campaign on an 8-m telescope, a sub-Neptune-sized planet could currently be imaged in the mid-IR [\citenum{Kasper2017,Kasper2019,Wagner2021}]. If the instrumental background can be further mitigated, then lower-mass planets$-$including those that are potentially habitable$-$can be directly imaged. Through subsequent (spectroscopic) observations, we can then study their atmospheric and orbital properties [e.g., \citenum{Snellen2015,Blunt2020}].

Here we review the state-of-the-art of mid-IR exoplanet imaging, some of the lessons learned so far, and some of the current work being done to continue progress toward imaging low-mass planets around nearby stars with these techniques. As a relatively new type of observation, much progress can be made by understanding and limiting the instrumental artifacts and sources of noise that were not known or not well-understood before the firsts on-sky tests. Now, with over 100 hours of data from a high-contrast mid-IR imager on the Very Large Telescope (VLT), and on-sky results beginning at the Large Binocular Telescope (LBT), a significant volume of data is available. This will enable us to understand and construct detailed systematic models, which can in turn be used to reduce the impact of some of the dominant sources of noise for mid-IR exoplanet imaging. 

\section{NEAR: New Earths in the Alpha Centauri Region}

The recently completed New Earths in the Alpha Centauri Region (NEAR) program explored several upgrades to the mid-IR imaging capabilities of the VLT. These enabled establishing the first sensitivity to habitable-zone Neptune-sized planets within the closest stellar system, $\alpha$ Centauri [\citenum{Kasper2019,Wagner2021}]. Some of the more significant instrumental upgrades included transferring the VLT's mid-IR camera, VISIR, to the unit telescope of the VLT equipped with the observatory's deformable secondary mirror (DSM)\footnote{The similar (deformable) optics within the LBT are frequently referred to as ``adaptive" secondary mirrors, or ASMs.}, which enables performing adaptive optics without additional warm optics [\citenum{Arsenault2017}]. VISIR was also equipped with a mid-IR optimized coronagraph [\citenum{Maire2020}] and a shaped-pupil mask specifically designed to limit the diffraction from the secondary star [\citenum{Ruane2015}]. 

\begin{figure} [ht]
   \begin{center}
   \begin{tabular}{c} 
   \includegraphics[height=8.5cm]{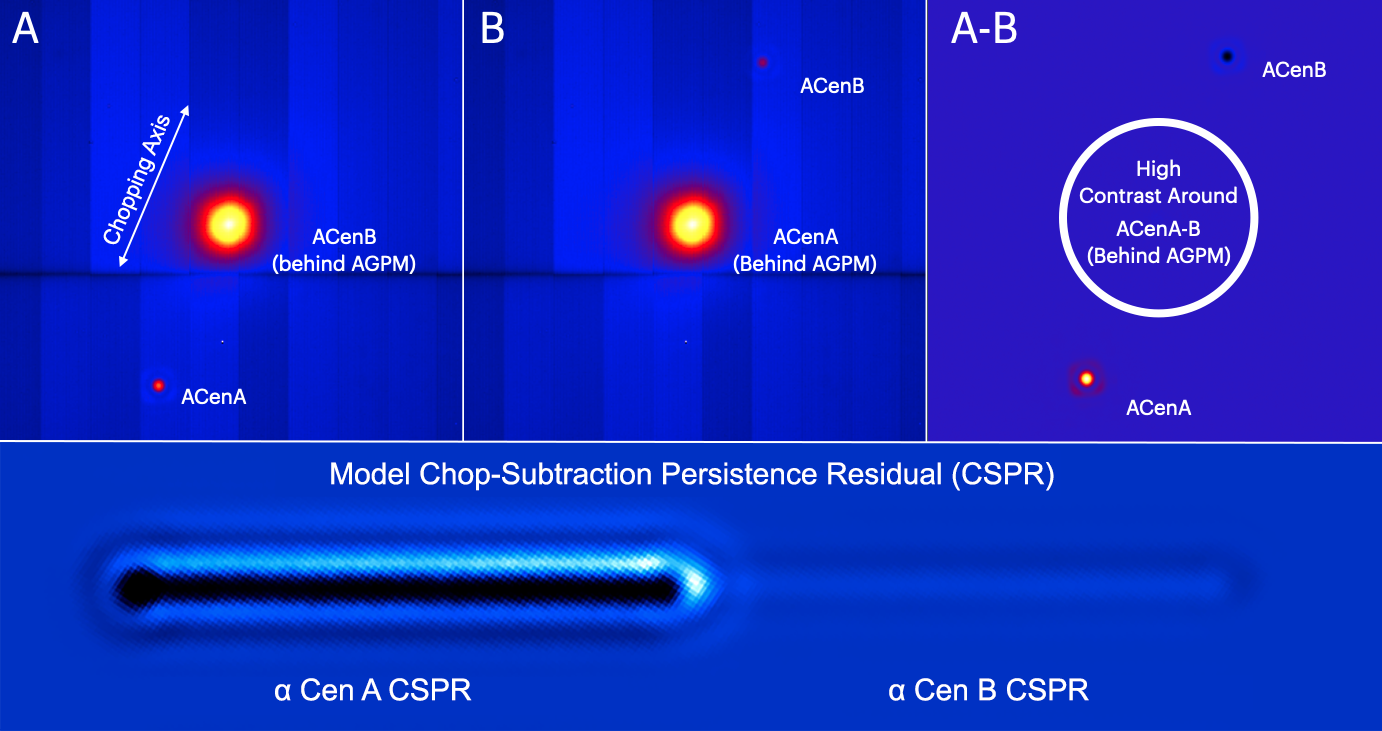}
   \end{tabular}
   \end{center}
   \caption[persistence] 
   { \label{fig:1} 
Top: an example of a chop-subtraction image pair from VLT/NEAR. Image B, following 0.05 sec after A, is subtracted to remove the detector bias, thermal background, AGPM glow, and excess low frequency noise (ELFN). Bottom: an example model of a chop-subtraction persistence residual (CSPR) of $\alpha$ Centauri AB. This model was generated with an offset of $\sim$2\% of a pixel, or $\sim$1 mas, of the star's position behind the coronagraph following each chop. The effects of the coronagraph, persistence non-linearity and temporal decay, and post-processing are not included in the model. The CSPR is too faint to be seen in the frames above, but is easily seen in longer integrations (see Fig. \ref{fig:2}).}

   \end{figure}

A crucial component of ground-based mid-IR imaging is the ability to measure and subtract the thermal background$-$including the contributions from the sky, telescope, and instrument$-$each of which have distinct temporal properties. The current generation of mid-IR detectors (specifically the Raytheon AQUARIUS Si:As arrays [\citenum{Ives2014}] within VISIR and NOMIC [\citenum{Hoffmann2014}], a similar camera installed at the LBT) also suffer from excess low frequency noise (ELFN) that increases exponentially toward lower frequencies [\citenum{Arrington1998}]. To mitigate the impact of this systematic noise feature, NEAR pioneered $\sim$10 Hz chopping using the DSM, which significantly reduced the background noise and enabled a dramatic improvement in the practicality of observing low-mass planets in realistic exposure times ($\lesssim$100 hr). Chopping also removes other static features, such as detector bias and the glow of the annular groove phase mask (AGPM) coronagraph  [\citenum{Absil2016}] that can be seen in Fig. \ref{fig:1}. 

However, chopping introduces other challenges. For coronagraphic observations, an obvious challenge is to maintain a precise positioning behind the mask following each chop cycle.\footnote{Centering is a particular problem of focal-plane coronagraphs. The newer pupil-plane apodized phase plate coronagraphs [\citenum{Snik2012,Doelman2021}] might be particularly well suited for chopping because they are tip/tilt invariant.} NEAR utilized the Quadrant Analysis of Coronagraphic Images for Tip-tilt Sensing (QACITS) algorithm [\citenum{Huby2016,Huby2017}], which results in $\sim$4 mas centering residuals ($\sim$0.1 pixels). This is also similar to the limit of precision of the DSM to re-position the star (with or without a coronagraph). Therefore, successive chops do not follow exactly the same trajectory. While these errors are randomly distributed and relatively small compared to the beam diameter, they introduce a major limitation to the current approach. This is due to a separate systematic of the AQUARIUS detectors$-$a persistence effect that becomes more pronounced for bright stars (the optimal targets) and that occurs even when the detector is not actively recording an exposure. This, combined with the random fractional-pixel displacements of the star with each chop cycle, results in a residual along the chopping axis that changes in form and intensity with each exposure pair. We refer to this feature as the chop-subtraction persistence residual, or CSPR (see Fig. \ref{fig:1}).

\begin{figure} [htpb]
   \begin{center}
   \begin{tabular}{c} 
   \includegraphics[height=7cm]{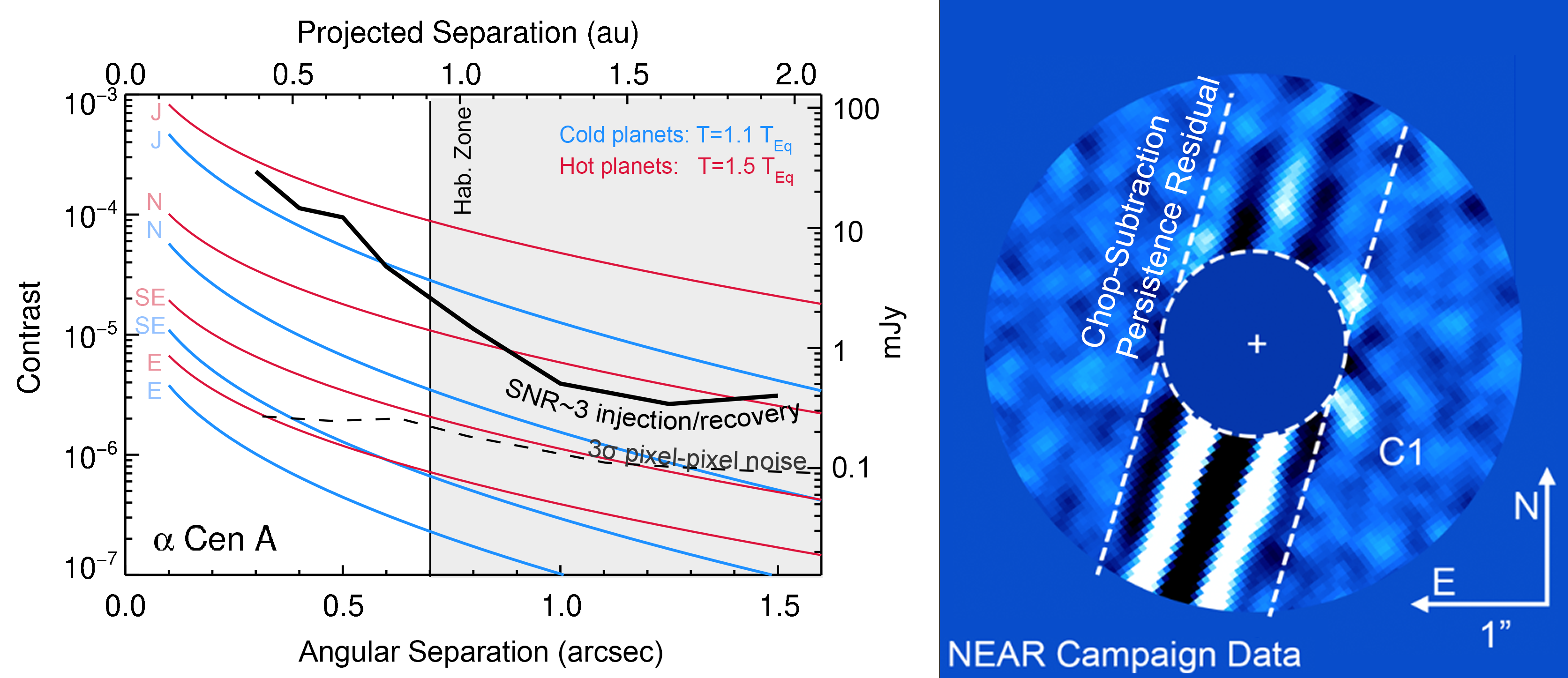}
   \end{tabular}
   \end{center}
   \caption[NEAR] 
   { \label{fig:2} 
Results of the VLT/NEAR experiment on $\alpha$ Centauri, reproduced from [\citenum{Wagner2021}]. Left: Contrast vs. angular separation of injected point sources recovered with a signal to noise ratio (SNR)~$\sim$~3 (black curve) and model planetary contrasts for hot and cold planets (red and blue curves). Right: image of the habitable zone of $\alpha$ Centauri, showing the CSPR feature and also a candidate signal around $\alpha$ Centauri A, labeled as C1 (see [\citenum{Wagner2021}] for more details).}

   \end{figure}

Given that typical exposure times of $\lesssim$0.1 sec are required to avoid saturation of the detector (due to thermal background) and to ensure correlated ELFN states, the CSPR becomes intractable to model and subtract in post-processing in a frame-by-frame manner. Furthermore, with NEAR's binary-chopping strategy, in which $\alpha$ Cen A and B are alternated in position behind the coronagraph, the persistence pattern rotates with the field of view and is not removed by angular differential imaging (ADI: \citenum{Marois2006}). Ultimately, this limited the sensitivity of the NEAR data and the regions most impacted by this feature were excluded from analysis (see Fig. \ref{fig:2}, reproduced from \citenum{Wagner2021}). This leads to two useful lessons for mid-IR exoplanet imaging with similar detectors: 1) improved chopping precision will lead to a less significant CSPR feature; and 2) chopping along a consistent path in the pupil-stabilized image will also enable the remaining CSPR to be removed in ADI-based post-processing. 

In addition to serving as a pathfinder experiment to explore the major challenges to mid-IR exoplanet imaging, the NEAR campaign succeeded in establishing the first sensitivity to warm sub-Neptune-sized planets within the habitable zone of $\alpha$ Centauri A [\citenum{Kasper2019, Wagner2021}]. This is a siginficant advancement for exoplanet imaging, which has been limited to studying planets of super-Jovian masses and on orbits of $\gtrsim$10 au. The sensitivity demonstrated by NEAR is also approaching levels sufficient to image rocky planets (R$\lesssim$1.7 R$_{\oplus}$) in the habitable zones of nearby stars$-$i.e., worlds that might support life. Now, a path is clear to implement the lessons brought by NEAR in existing and future mid-IR instruments to ultimately enable us to achieve the goal of finding a second Earth.

\section{Building on the Lessons of NEAR with LBTI/NOMIC}

 \begin{figure} [ht]
   \begin{center}
   \begin{tabular}{c} 
   \includegraphics[height=9cm]{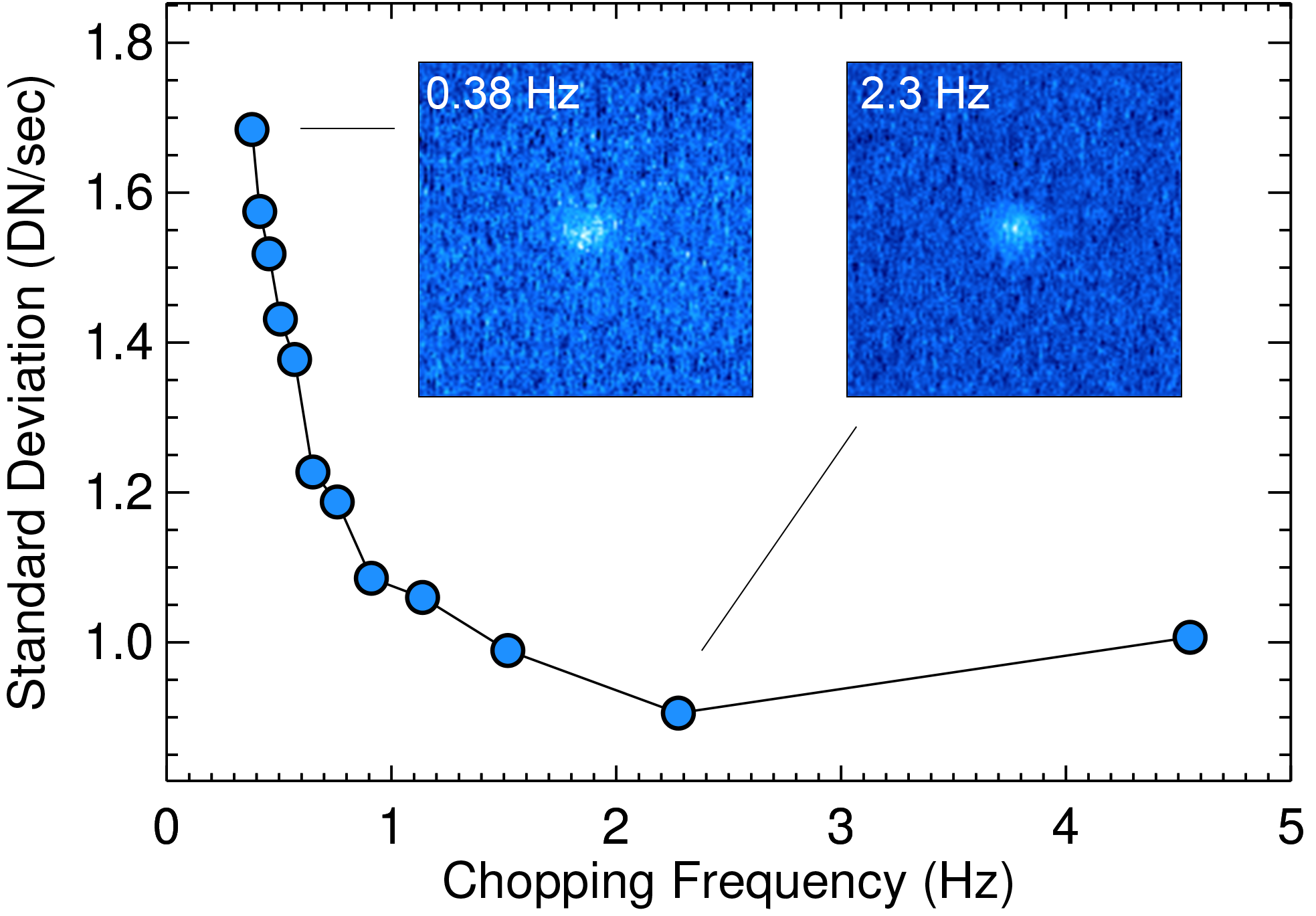}
   \end{tabular}
   \end{center}
   \caption[elfntest] 
   { \label{fig:3} 
Pixel-to-pixel noise as a function of the chopping half-cycle frequency with LBTI/NOMIC. Similar to results from the NEAR experiment with VLT/VISIR, we find a rapidly decreasing level of noise with increasing chopping frequency up to 2.3 Hz. The 5$-$10\% increase in noise between 2.3 Hz and 4.55 Hz is a feature not seen previously with VISIR, but that has been seen repeatedly in multiple LBTI/NOMIC datasets. }
   \end{figure}

In the Northern hemisphere, telescopes such as the LBT in Arizona, those situated on Mauna Kea, and several other facilities present an opportunity for continued progress to improve mid-IR exoplanet imaging instrumentation. The LBT's location on the high-altitude inland peak of Mount Graham experiences average winter temperatures that are 15$-$20$^{\circ}$ C colder than the coastal sites of other large telescopes, while providing similar precipitable water vapor conditions [\citenum{Turchi2019}]. This results in a thermal background that is $\sim$25$-$30\% lower than other sites.  Furthermore, with its dual apertures, the LBT offers an additional improvement of a factor of at least $\sqrt{2}$ for a given amount of time (a larger factor can be gained by Fizeau imaging, see below). Most of the necessary exoplanet imaging hardware exists within the framework of the Large Binocular Telescope Interferometer (LBTI: \citenum{Hinz2016, Ertel2020}), as was recently demonstrated by [\citenum{Stone2018,Wagner2019,Wagner2020}]. LBTI's bent-Gregorian position directly on the telescope mount is fixed with respect to the telescope pupil, which also leads to a relatively stable thermal background. Each side of the telescope contains a low-background adaptive secondary mirror (ASM) that can be simultaneously imaged with NOMIC [\citenum{Hoffmann2014}], whose AQUARIUS detector is nearly identical to that within VISIR. 

LBTI is a focal plane interferometer that operates by overlapping the individual point spread functions of each 8.4-m mirror of the LBT on a single detector. 23-m nulling and Fizeau imaging is achieved using cryogenic pathlength correctors (pupil-mirrors) operating at kHz frequencies to correct tip, tilt, and piston abberations in order to keep the beams overlapped and coherent  [\citenum{Ertel2018, Spalding2019, Ertel2020b, Ertel2020}]. LBTI can also be used in a classical (incoherent) imaging mode, in which the beams are displaced to form two images of a single star on the detector [e.g., \citenum{Skemer2014, Stone2018, Wagner2019, Wagner2020}]. We have developed a method using LBTI's pupil-mirrors for chopping when observing in an incoherent mode. Chopping with the pupil-mirrors does not introduce additional background nor does it impact adaptive optics operations. The control loop required for chopping is also much slower than the $\sim$kHz rates used for interferometry. Chopping with the pupil-mirrors takes place in the pupil-stabilized frame$-$enabling the remaining CSPR to be removed with ADI-based post-processing. As of mid-2021 we have commissioned up to $\sim$4.5 Hz chopping with the pupil-mirrors, with faster rates being limited by NOMIC's readout capabilities. Similar to NEAR, faster chopping (up to rates of $\sim$2 Hz) results in exponentially lower background noise (Fig. \ref{fig:3}). However, unlike NEAR, most on-sky LBTI/NOMIC chopping datasets show an optimal chopping rate around $\sim$2 Hz, with the single higher rate showing a $\sim$10\% increase in noise. The reason for this is not yet understood, but is possibly related to the smaller number of coadded (0.04 second) frames per half-cycle position.


 \begin{figure} [ht]
   \begin{center}
   \begin{tabular}{c} 
   \includegraphics[height=11cm]{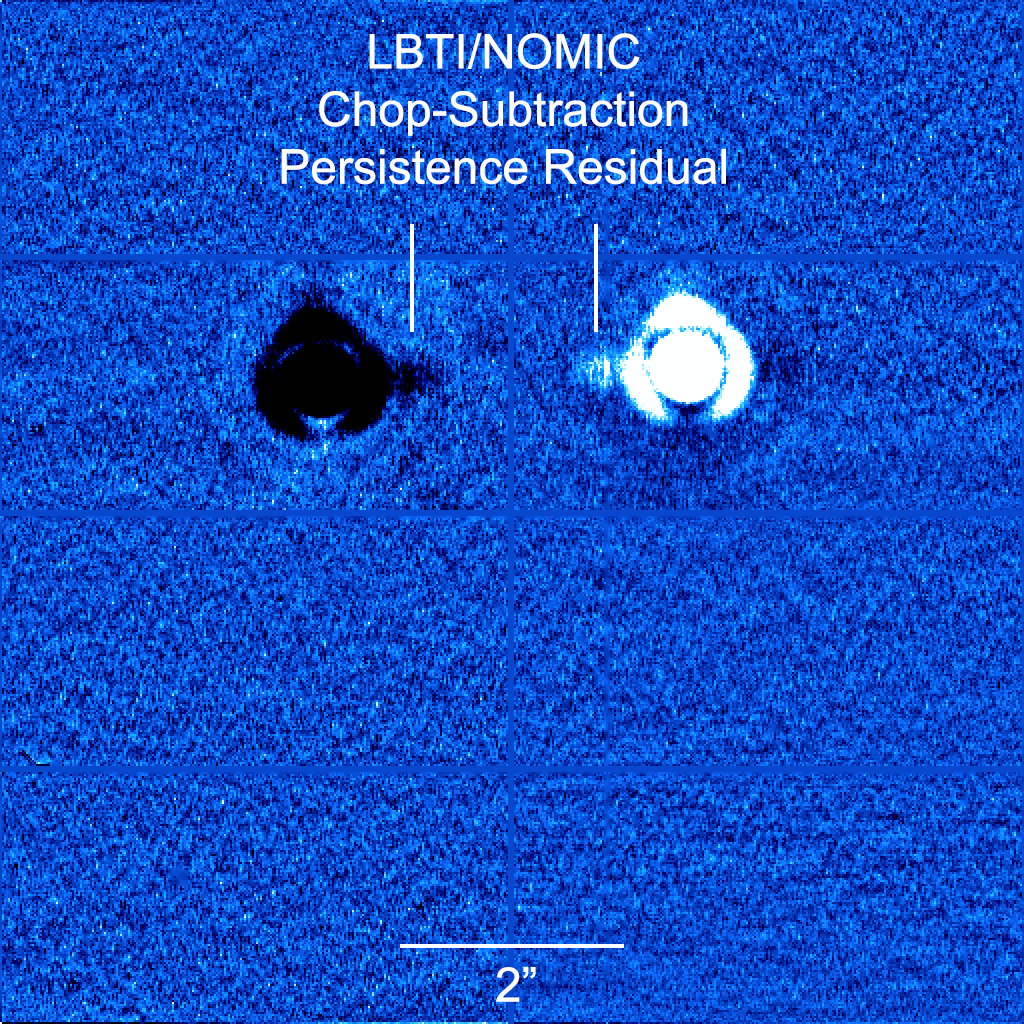}
   \end{tabular}
   \end{center}
   \caption[Sirius_chop] 
   { \label{fig:4} 
An example of a 512x512 chop-subtracted image pair of Sirius from NOMIC. A CSPR feature similar to that in the NEAR data is easily identified adjacent to each PSF along the chopping axis. This artifact is stable with the telescope pupil and is subtracted with ADI-based image processing. Once double-sided operations resume at the LBT, the bottom half of the image plane will be used to simultaneously observe with the second aperture.}
   \end{figure} 

The LBT offers another avenue for improvement with its dual 8.4-m apertures. Unfortunately, its most sensitive imaging mode (in principle), Fizeau interferometry, cannot currently be combined with chopping due to the fringe-tracker's narrow field of view, and in this case the ELFN quickly precludes any gain in sensitivity. Overlapping the two beams incoherently results in an additional $\sqrt{2}$ gain in sensitivity [\citenum{Skemer2016}], but loses discriminating power in the independent speckle patterns. By installing new cold field-stops, we plan to enable background-limited imaging of the two beams simultaneously and separately on the detector (see Fig. \ref{fig:4}). This will enable working at smaller separations via identification of true sources within the independent speckle patterns (speckle-limited imaging). The cold field-stops have been designed and fabricated specifically for this project and will be installed over Summer 2021. Once both adaptive secondary mirrors are operational (planned for late 2021), double-sided observations will enable imaging habitable-zone sub-Neptune-sized planets in as little as $\sim$6 nights.

 \begin{figure} [ht]
   \begin{center}
   \begin{tabular}{c} 
   \includegraphics[height=7cm]{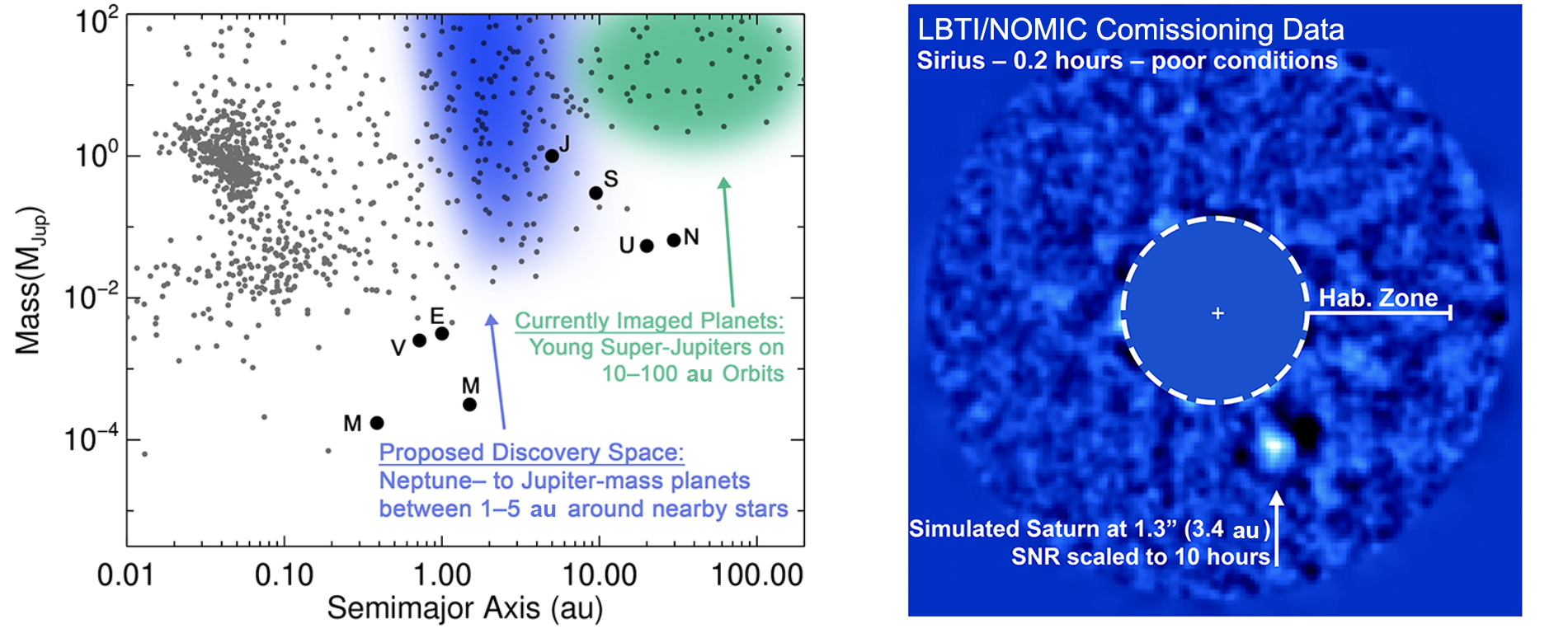}
   \end{tabular}
   \end{center}
   \caption[discovery-space-au] 
   { \label{fig:5} 
Left: masses vs. orbital semi-major axes of known exoplanets from all detection methods. The planets detectable with mid-IR high-contrast imaging in less than six nights with a single 8.4-m telescope are shown in blue$-$representing a distinct and complementary discovery space to currently imaged super-Jupiters. Right: an example of a simulated Saturn-sized planet in the habitable zone of Sirius. Such planets would be readily detectable in 1$-$2 nights of observations.}

   \end{figure}

 \section{Exoplanet Science With High-Contrast Mid-IR Imaging}

The primary scientific benefit of mid-IR exoplanet imaging (including exo-Earth imaging capabilities: \citenum{Wagner2021,Brandl2021,Kasper2021}) will come with the next generation of extremely large telescopes (ELTs). However, it is useful to briefly consider the science that will be enabled by the current generation of experimental instruments on 8-m telescopes. With high-contrast mid-IR imaging capabilities now in both hemispheres, and others in preparation [e.g., \citenum{Blain2018, Morzinski2020}], it will be possible to conduct a complete survey for habitable-zone giant planets around nearby stars. While reflected light imaging [e.g., \citenum{Roman}] will also enable the study of many of these same planets, its combination with mid-IR imaging is needed to break the degeneracy between radius and albedo. Indeed, such observations have already started to place limits on the properties of planets within a handful of systems [e.g., \citenum{Mawet2019, Pathak2021, Viswanath2021, Wagner2021}].

The frequency of directly imaged giant planets is increasing toward smaller masses and smaller separations [\citenum{Wagner2019b,Nielsen2019,Vigan2020}]. Furthermore, the peak in the occurrence of giant planets exists at $\lesssim$10 au [\citenum{Fernandes2019,Fulton2021}]. Therefore, the probability of discovering a temperate giant planet around one of the nearby stars is relatively high. To quantify the discovery space of a complete mid-IR exoplanet imaging survey of nearby stars (those within $\sim$15 pc), we simulated the $N$-band ($\sim$11 $\mu$m) contrast of habitable-zone giant planets following [\citenum{Wagner2021}] and assuming planetary temperatures determined by radiative thermal equilibrium. This resulted in a list of ten stars around which Jupiter$-$ and Saturn-sized planets between 1$-$5 au will result in a $\geq$5-$\sigma$ detection in a single night of observations.  Current population estimates suggest that $\sim$10$-$40\% of stars host a 0.3$-$13 Jupiter-mass planet between 1$-$5 au [\citenum{Fernandes2019}], which translates into an expected population of $\sim$1$-$4 nearby habitable-zone giant planets.\footnote{While all of these would be detectable over multiple epochs, a single observation may miss some highly-inclined planets that are in close projected separation to their stars.}

Since the sensitivity scales roughly with $\sqrt{t}$, it will be possible to push even deeper by observing each target for multiple nights and with multiple instruments.  Based on commissioning data, six nights of observations with both sides of the LBTI will be sufficient for a 3-$\sigma$ detection of a sub-Neptune-sized planet at $\sim$2$-$3 au around Sirius.  Such worlds would occupy a new range of parameter space for imaged exoplanets (see Fig. \ref{fig:5}). As no sub-Jovian exoplanets have been directly imaged\footnote{Transit and high-resolution spectroscopy, however, can be an effective means for studying close-in planets.}, this would open up a new class of planets for detailed atmospheric and orbital characterization. This is the motivation behind our on-going LBTI Exploratory Survey for Super-Earths Orbitng Nearby Stars (LESSONS). These observations will also demonstrate the ability to combine data from multiple nights and will test new data reduction methods for handling orbital motion [e.g., \citenum{Males2015,LeCoroller2020}]. Finally, the upcoming ELTs will be able to efficiently image super-Earths at radiative equilibrium temperatures around several nearby stars, and possibly Earth-sized planets within the closest system, $\alpha$ Centauri [\citenum{Brandl2021,Kasper2021,Wagner2021}]. NEAR and LESSONS will serve as pathfinders for these groundbreaking instruments [e.g., \citenum{Fitzgerald2019,Brandl2021}].

\section{Summary}

We have reviewed the current status of mid-IR ($\sim$10 $\mu$m) exoplanet imaging instrumentation and presented results on recent progress to expand these capabilities. The most sensitive observations to date were taken by the NEAR campaign, which established sensitivity to warm Neptune-sized planets around $\alpha$ Centauri A. As a first of its kind long-stare mid-IR campaign for exoplanet imaging, NEAR also revealed a number of promising avenues to improve future instrumentation. The main results of NEAR can be found in [\citenum{Kasper2017,Kasper2019,Wagner2021}]. Here, we focused on a primary challenge to NEAR's strategy: persistence combining with chop-subtraction to create a complicated systematic artifact. We discussed the properties of this feature and how it can be mitigated in the future. We introduced a new experiment, LESSONS, through which we aim to build on the results of NEAR to continue pushing sensitivity toward imaging low-mass planets. Finally, we discussed how mid-IR exoplanet imaging will enable new science. It is now possible to discover and characterize exoplanets that are roughly an order of magnitude lower in mass than those that have so far been directly imaged. These planets will also be at orbital separations that are much closer-in than currently imaged exoplanets$-$enabling new studies across a broad range of physical parameters. Ultimately, the sample capable of being targeted with current telescopes is limited by the number of bright nearby stars. In the future, the ELTs will enable expanding the list of target stars and will enable imaging potentially Earth-like planets within a handful of nearby systems.

\acknowledgments 
 
NEAR was made possible by contributions from the Breakthrough Watch program, as
well as contributions from the European Southern Observatory, including director’s
discretionary time. Breakthrough Watch is managed by the Breakthrough Initiatives,
sponsored by the Breakthrough Prize Foundation. The results reported herein benefited from collaborations and/or information exchange within NASA's Nexus for Exoplanet System Science (NExSS) research coordination network sponsored by NASA's Science Mission Directorate. K.W. acknowledges support from NASA through the NASA Hubble Fellowship grant HST-HF2-51472.001-A awarded by the Space Telescope Science Institute, which is operated by the Association of Universities for Research in Astronomy, Incorporated, under NASA contract NAS5-26555.

\bibliography{main} 

\begin{thebibliography}{10}

\bibitem{Charbonneau2009}
{Charbonneau}, D., {Berta}, Z.~K., {Irwin}, J., {Burke}, C.~J., {Nutzman}, P.,
  {Buchhave}, L.~A., {Lovis}, C., {Bonfils}, X., {Latham}, D.~W., {Udry}, S.,
  {Murray-Clay}, R.~A., {Holman}, M.~J., {Falco}, E.~E., {Winn}, J.~N.,
  {Queloz}, D., {Pepe}, F., {Mayor}, M., {Delfosse}, X., and {Forveille}, T.,
  ``{A super-Earth transiting a nearby low-mass star},'' {\em Nature}~{\bf
  462},  891--894 (Dec. 2009).

\bibitem{Jenkins2015}
{Jenkins}, J.~M., {Twicken}, J.~D., {Batalha}, N.~M., {Caldwell}, D.~A.,
  {Cochran}, W.~D., {Endl}, M., {Latham}, D.~W., {Esquerdo}, G.~A., {Seader},
  S., {Bieryla}, A., {Petigura}, E., {Ciardi}, D.~R., {Marcy}, G.~W.,
  {Isaacson}, H., {Huber}, D., {Rowe}, J.~F., {Torres}, G., {Bryson}, S.~T.,
  {Buchhave}, L., {Ramirez}, I., {Wolfgang}, A., {Li}, J., {Campbell}, J.~R.,
  {Tenenbaum}, P., {Sanderfer}, D., {Henze}, C.~E., {Catanzarite}, J.~H.,
  {Gilliland}, R.~L., and {Borucki}, W.~J., ``{Discovery and Validation of
  Kepler-452b: A 1.6 R $_{\oplus}$ Super Earth Exoplanet in the Habitable Zone
  of a G2 Star},'' {\em AJ}~{\bf 150},  56 (Aug. 2015).

\bibitem{Berta2015}
{Berta-Thompson}, Z.~K., {Irwin}, J., {Charbonneau}, D., {Newton}, E.~R.,
  {Dittmann}, J.~A., {Astudillo-Defru}, N., {Bonfils}, X., {Gillon}, M.,
  {Jehin}, E., {Stark}, A.~A., {Stalder}, B., {Bouchy}, F., {Delfosse}, X.,
  {Forveille}, T., {Lovis}, C., {Mayor}, M., {Neves}, V., {Pepe}, F., {Santos},
  N.~C., {Udry}, S., and {W{\"u}nsche}, A., ``{A rocky planet transiting a
  nearby low-mass star},'' {\em Nature}~{\bf 527},  204--207 (Nov. 2015).

\bibitem{Dittmann2015}
{Dittmann}, J.~A., {Irwin}, J.~M., {Charbonneau}, D., {Bonfils}, X.,
  {Astudillo-Defru}, N., {Haywood}, R.~D., {Berta-Thompson}, Z.~K., {Newton},
  E.~R., {Rodriguez}, J.~E., {Winters}, J.~G., {Tan}, T.-G., {Almenara}, J.-M.,
  {Bouchy}, F., {Delfosse}, X., {Forveille}, T., {Lovis}, C., {Murgas}, F.,
  {Pepe}, F., {Santos}, N.~C., {Udry}, S., {W{\"u}nsche}, A., {Esquerdo},
  G.~A., {Latham}, D.~W., and {Dressing}, C.~D., ``{A temperate rocky
  super-Earth transiting a nearby cool star},'' {\em Nature}~{\bf 544},
  333--336 (Apr. 2017).

\bibitem{Anglada2016}
{Anglada-Escud{\'e}}, G., {Amado}, P.~J., {Barnes}, J., {Berdi{\~n}as}, Z.~M.,
  {Butler}, R.~P., {Coleman}, G. A.~L., {de La Cueva}, I., {Dreizler}, S.,
  {Endl}, M., {Giesers}, B., {Jeffers}, S.~V., {Jenkins}, J.~S., {Jones}, H.
  R.~A., {Kiraga}, M., {K{\"u}rster}, M., {L{\'o}pez-Gonz{\'a}lez}, M.~J.,
  {Marvin}, C.~J., {Morales}, N., {Morin}, J., {Nelson}, R.~P., {Ortiz}, J.~L.,
  {Ofir}, A., {Paardekooper}, S.-J., {Reiners}, A., {Rodr{\'\i}guez}, E.,
  {Rodr{\'\i}guez-L{\'o}pez}, C., {Sarmiento}, L.~F., {Strachan}, J.~P.,
  {Tsapras}, Y., {Tuomi}, M., and {Zechmeister}, M., ``{A terrestrial planet
  candidate in a temperate orbit around Proxima Centauri},'' {\em Nature}~{\bf
  536},  437--440 (Aug. 2016).

\bibitem{Gillon2017}
{Gillon}, M., {Triaud}, A. H.~M.~J., {Demory}, B.-O., {Jehin}, E., {Agol}, E.,
  {Deck}, K.~M., {Lederer}, S.~M., {de Wit}, J., {Burdanov}, A., {Ingalls},
  J.~G., {Bolmont}, E., {Leconte}, J., {Raymond}, S.~N., {Selsis}, F.,
  {Turbet}, M., {Barkaoui}, K., {Burgasser}, A., {Burleigh}, M.~R., {Carey},
  S.~J., {Chaushev}, A., {Copperwheat}, C.~M., {Delrez}, L., {Fernandes},
  C.~S., {Holdsworth}, D.~L., {Kotze}, E.~J., {Van Grootel}, V., {Almleaky},
  Y., {Benkhaldoun}, Z., {Magain}, P., and {Queloz}, D., ``{Seven temperate
  terrestrial planets around the nearby ultracool dwarf star TRAPPIST-1},''
  {\em Nature}~{\bf 542},  456--460 (Feb. 2017).

\bibitem{Marois2010}
{Marois}, C., {Zuckerman}, B., {Konopacky}, Q.~M., {Macintosh}, B., and
  {Barman}, T., ``{Images of a fourth planet orbiting HR 8799},'' {\em
  Nature}~{\bf 468},  1080--1083 (Dec. 2010).

\bibitem{Lagrange2010}
{Lagrange}, A.~M., {Bonnefoy}, M., {Chauvin}, G., {Apai}, D., {Ehrenreich}, D.,
  {Boccaletti}, A., {Gratadour}, D., {Rouan}, D., {Mouillet}, D., {Lacour}, S.,
  and {Kasper}, M., ``{A Giant Planet Imaged in the Disk of the Young Star
  {\ensuremath{\beta}} Pictoris},'' {\em Science}~{\bf 329},  57 (July 2010).

\bibitem{Rameau2013}
{Rameau}, J., {Chauvin}, G., {Lagrange}, A.~M., {Meshkat}, T., {Boccaletti},
  A., {Quanz}, S.~P., {Currie}, T., {Mawet}, D., {Girard}, J.~H., {Bonnefoy},
  M., and {Kenworthy}, M., ``{Confirmation of the Planet around HD 95086 by
  Direct Imaging},'' {\em ApJl}~{\bf 779},  L26 (Dec. 2013).

\bibitem{Macintosh2015}
{Macintosh}, B., {Graham}, J.~R., {Barman}, T., {De Rosa}, R.~J., {Konopacky},
  Q., {Marley}, M.~S., {Marois}, C., {Nielsen}, E.~L., {Pueyo}, L., {Rajan},
  A., {Rameau}, J., {Saumon}, D., {Wang}, J.~J., {Patience}, J., {Ammons}, M.,
  {Arriaga}, P., {Artigau}, E., {Beckwith}, S., {Brewster}, J., {Bruzzone}, S.,
  {Bulger}, J., {Burningham}, B., {Burrows}, A.~S., {Chen}, C., {Chiang}, E.,
  {Chilcote}, J.~K., {Dawson}, R.~I., {Dong}, R., {Doyon}, R., {Draper}, Z.~H.,
  {Duch{\^e}ne}, G., {Esposito}, T.~M., {Fabrycky}, D., {Fitzgerald}, M.~P.,
  {Follette}, K.~B., {Fortney}, J.~J., {Gerard}, B., {Goodsell}, S.,
  {Greenbaum}, A.~Z., {Hibon}, P., {Hinkley}, S., {Cotten}, T.~H., {Hung},
  L.~W., {Ingraham}, P., {Johnson-Groh}, M., {Kalas}, P., {Lafreniere}, D.,
  {Larkin}, J.~E., {Lee}, J., {Line}, M., {Long}, D., {Maire}, J., {Marchis},
  F., {Matthews}, B.~C., {Max}, C.~E., {Metchev}, S., {Millar-Blanchaer},
  M.~A., {Mittal}, T., {Morley}, C.~V., {Morzinski}, K.~M., {Murray-Clay}, R.,
  {Oppenheimer}, R., {Palmer}, D.~W., {Patel}, R., {Perrin}, M.~D., {Poyneer},
  L.~A., {Rafikov}, R.~R., {Rantakyr{\"o}}, F.~T., {Rice}, E.~L., {Rojo}, P.,
  {Rudy}, A.~R., {Ruffio}, J.~B., {Ruiz}, M.~T., {Sadakuni}, N., {Saddlemyer},
  L., {Salama}, M., {Savransky}, D., {Schneider}, A.~C., {Sivaramakrishnan},
  A., {Song}, I., {Soummer}, R., {Thomas}, S., {Vasisht}, G., {Wallace}, J.~K.,
  {Ward-Duong}, K., {Wiktorowicz}, S.~J., {Wolff}, S.~G., and {Zuckerman}, B.,
  ``{Discovery and spectroscopy of the young jovian planet 51 Eri b with the
  Gemini Planet Imager},'' {\em Science}~{\bf 350},  64--67 (Oct. 2015).

\bibitem{Chauvin2017}
{Chauvin}, G., {Desidera}, S., {Lagrange}, A.~M., {Vigan}, A., {Gratton}, R.,
  {Langlois}, M., {Bonnefoy}, M., {Beuzit}, J.~L., {Feldt}, M., {Mouillet}, D.,
  {Meyer}, M., {Cheetham}, A., {Biller}, B., {Boccaletti}, A., {D'Orazi}, V.,
  {Galicher}, R., {Hagelberg}, J., {Maire}, A.~L., {Mesa}, D., {Olofsson}, J.,
  {Samland}, M., {Schmidt}, T.~O.~B., {Sissa}, E., {Bonavita}, M., {Charnay},
  B., {Cudel}, M., {Daemgen}, S., {Delorme}, P., {Janin-Potiron}, P., {Janson},
  M., {Keppler}, M., {Le Coroller}, H., {Ligi}, R., {Marleau}, G.~D.,
  {Messina}, S., {Molli{\`e}re}, P., {Mordasini}, C., {M{\"u}ller}, A.,
  {Peretti}, S., {Perrot}, C., {Rodet}, L., {Rouan}, D., {Zurlo}, A.,
  {Dominik}, C., {Henning}, T., {Menard}, F., {Schmid}, H.~M., {Turatto}, M.,
  {Udry}, S., {Vakili}, F., {Abe}, L., {Antichi}, J., {Baruffolo}, A.,
  {Baudoz}, P., {Baudrand}, J., {Blanchard}, P., {Bazzon}, A., {Buey}, T.,
  {Carbillet}, M., {Carle}, M., {Charton}, J., {Cascone}, E., {Claudi}, R.,
  {Costille}, A., {Deboulbe}, A., {De Caprio}, V., {Dohlen}, K., {Fantinel},
  D., {Feautrier}, P., {Fusco}, T., {Gigan}, P., {Giro}, E., {Gisler}, D.,
  {Gluck}, L., {Hubin}, N., {Hugot}, E., {Jaquet}, M., {Kasper}, M., {Madec},
  F., {Magnard}, Y., {Martinez}, P., {Maurel}, D., {Le Mignant}, D.,
  {M{\"o}ller-Nilsson}, O., {Llored}, M., {Moulin}, T., {Orign{\'e}}, A.,
  {Pavlov}, A., {Perret}, D., {Petit}, C., {Pragt}, J., {Puget}, P., {Rabou},
  P., {Ramos}, J., {Rigal}, R., {Rochat}, S., {Roelfsema}, R., {Rousset}, G.,
  {Roux}, A., {Salasnich}, B., {Sauvage}, J.~F., {Sevin}, A., {Soenke}, C.,
  {Stadler}, E., {Suarez}, M., {Weber}, L., {Wildi}, F., {Antoniucci}, S.,
  {Augereau}, J.~C., {Baudino}, J.~L., {Brandner}, W., {Engler}, N., {Girard},
  J., {Gry}, C., {Kral}, Q., {Kopytova}, T., {Lagadec}, E., {Milli}, J.,
  {Moutou}, C., {Schlieder}, J., {Szul{\'a}gyi}, J., {Thalmann}, C., and
  {Wahhaj}, Z., ``{Discovery of a warm, dusty giant planet around HIP 65426},''
  {\em A\&A}~{\bf 605},  L9 (Sept. 2017).

\bibitem{Keppler2018}
{Keppler}, M., {Benisty}, M., {M{\"u}ller}, A., {Henning}, T., {van Boekel},
  R., {Cantalloube}, F., {Ginski}, C., {van Holstein}, R.~G., {Maire}, A.~L.,
  {Pohl}, A., {Samland}, M., {Avenhaus}, H., {Baudino}, J.~L., {Boccaletti},
  A., {de Boer}, J., {Bonnefoy}, M., {Chauvin}, G., {Desidera}, S., {Langlois},
  M., {Lazzoni}, C., {Marleau}, G.~D., {Mordasini}, C., {Pawellek}, N.,
  {Stolker}, T., {Vigan}, A., {Zurlo}, A., {Birnstiel}, T., {Brandner}, W.,
  {Feldt}, M., {Flock}, M., {Girard}, J., {Gratton}, R., {Hagelberg}, J.,
  {Isella}, A., {Janson}, M., {Juhasz}, A., {Kemmer}, J., {Kral}, Q.,
  {Lagrange}, A.~M., {Launhardt}, R., {Matter}, A., {M{\'e}nard}, F., {Milli},
  J., {Molli{\`e}re}, P., {Olofsson}, J., {P{\'e}rez}, L., {Pinilla}, P.,
  {Pinte}, C., {Quanz}, S.~P., {Schmidt}, T., {Udry}, S., {Wahhaj}, Z.,
  {Williams}, J.~P., {Buenzli}, E., {Cudel}, M., {Dominik}, C., {Galicher}, R.,
  {Kasper}, M., {Lannier}, J., {Mesa}, D., {Mouillet}, D., {Peretti}, S.,
  {Perrot}, C., {Salter}, G., {Sissa}, E., {Wildi}, F., {Abe}, L., {Antichi},
  J., {Augereau}, J.~C., {Baruffolo}, A., {Baudoz}, P., {Bazzon}, A., {Beuzit},
  J.~L., {Blanchard}, P., {Brems}, S.~S., {Buey}, T., {De Caprio}, V.,
  {Carbillet}, M., {Carle}, M., {Cascone}, E., {Cheetham}, A., {Claudi}, R.,
  {Costille}, A., {Delboulb{\'e}}, A., {Dohlen}, K., {Fantinel}, D.,
  {Feautrier}, P., {Fusco}, T., {Giro}, E., {Gluck}, L., {Gry}, C., {Hubin},
  N., {Hugot}, E., {Jaquet}, M., {Le Mignant}, D., {Llored}, M., {Madec}, F.,
  {Magnard}, Y., {Martinez}, P., {Maurel}, D., {Meyer}, M.,
  {M{\"o}ller-Nilsson}, O., {Moulin}, T., {Mugnier}, L., {Orign{\'e}}, A.,
  {Pavlov}, A., {Perret}, D., {Petit}, C., {Pragt}, J., {Puget}, P., {Rabou},
  P., {Ramos}, J., {Rigal}, F., {Rochat}, S., {Roelfsema}, R., {Rousset}, G.,
  {Roux}, A., {Salasnich}, B., {Sauvage}, J.~F., {Sevin}, A., {Soenke}, C.,
  {Stadler}, E., {Suarez}, M., {Turatto}, M., and {Weber}, L., ``{Discovery of
  a planetary-mass companion within the gap of the transition disk around PDS
  70},'' {\em A\&A}~{\bf 617},  A44 (Sept. 2018).

\bibitem{Bohn2020}
{Bohn}, A.~J., {Kenworthy}, M.~A., {Ginski}, C., {Rieder}, S., {Mamajek},
  E.~E., {Meshkat}, T., {Pecaut}, M.~J., {Reggiani}, M., {de Boer}, J.,
  {Keller}, C.~U., {Snik}, F., and {Southworth}, J., ``{Two Directly Imaged,
  Wide-orbit Giant Planets around the Young, Solar Analog TYC 8998-760-1},''
  {\em ApJl}~{\bf 898},  L16 (July 2020).

\bibitem{Kasper2017}
{Kasper}, M., {Arsenault}, R., {K{\"a}ufl}, H.~U., {Jakob}, G., {Fuenteseca},
  E., {Riquelme}, M., {Siebenmorgen}, R., {Sterzik}, M., {Zins}, G.,
  {Ageorges}, N., {Gutruf}, S., {Reutlinger}, A., {Kampf}, D., {Absil}, O.,
  {Carlomagno}, B., {Guyon}, O., {Klupar}, P., {Mawet}, D., {Ruane}, G.,
  {Karlsson}, M., {Pantin}, E., and {Dohlen}, K., ``{NEAR: Low-mass Planets in
  {\ensuremath{\alpha}} Cen with VISIR},'' {\em The Messenger}~{\bf 169},
  16--20 (Sept. 2017).

\bibitem{Kasper2019}
{Kasper}, M., {Arsenault}, R., {K{\"a}ufl}, U., {Jakob}, G., {Leveratto}, S.,
  {Zins}, G., {Pantin}, E., {Duhoux}, P., {Riquelme}, M., {Kirchbauer}, J.~P.,
  {Kolb}, J., {Pathak}, P., {Siebenmorgen}, R., {Soenke}, C., {Fuenteseca}, E.,
  {Sterzik}, M., {Ageorges}, N., {Gutruf}, S., {Kampf}, D., {Reutlinger}, A.,
  {Absil}, O., {Delacroix}, C., {Maire}, A.~L., {Huby}, E., {Guyon}, O.,
  {Klupar}, P., {Mawet}, D., {Ruane}, G., {Karlsson}, M., {Dohlen}, K.,
  {Vigan}, A., {N'Diaye}, M., {Quanz}, S., and {Carlotti}, A., ``{NEAR: First
  Results from the Search for Low-Mass Planets in {\ensuremath{\alpha}} Cen},''
  {\em The Messenger}~{\bf 178},  5--9 (Dec. 2019).

\bibitem{Wagner2021}
{Wagner}, K., {Boehle}, A., {Pathak}, P., {Kasper}, M., {Arsenault}, R.,
  {Jakob}, G., {K{\"a}ufl}, U., {Leveratto}, S., {Maire}, A.~L., {Pantin}, E.,
  {Siebenmorgen}, R., {Zins}, G., {Absil}, O., {Ageorges}, N., {Apai}, D.,
  {Carlotti}, A., {Choquet}, {\'E}., {Delacroix}, C., {Dohlen}, K., {Duhoux},
  P., {Forsberg}, P., {Fuenteseca}, E., {Gutruf}, S., {Guyon}, O., {Huby}, E.,
  {Kampf}, D., {Karlsson}, M., {Kervella}, P., {Kirchbauer}, J.~P., {Klupar},
  P., {Kolb}, J., {Mawet}, D., {N'Diaye}, M., {Orban de Xivry}, G., {Quanz},
  S.~P., {Reutlinger}, A., {Ruane}, G., {Riquelme}, M., {Soenke}, C.,
  {Sterzik}, M., {Vigan}, A., and {de Zeeuw}, T., ``{Imaging low-mass planets
  within the habitable zone of {\ensuremath{\alpha}} Centauri},'' {\em Nature
  Communications}~{\bf 12},  922 (Jan. 2021).

\bibitem{Snellen2015}
{Snellen}, I., {de Kok}, R., {Birkby}, J.~L., {Brandl}, B., {Brogi}, M.,
  {Keller}, C., {Kenworthy}, M., {Schwarz}, H., and {Stuik}, R., ``{Combining
  high-dispersion spectroscopy with high contrast imaging: Probing rocky
  planets around our nearest neighbors},'' {\em A\&A}~{\bf 576},  A59 (Apr.
  2015).

\bibitem{Blunt2020}
{Blunt}, S., {Wang}, J.~J., {Angelo}, I., {Ngo}, H., {Cody}, D., {De Rosa},
  R.~J., {Graham}, J.~R., {Hirsch}, L., {Nagpal}, V., {Nielsen}, E.~L.,
  {Pearce}, L., {Rice}, M., and {Tejada}, R., ``{orbitize!: A Comprehensive
  Orbit-fitting Software Package for the High-contrast Imaging Community},''
  {\em AJ}~{\bf 159},  89 (Mar. 2020).

\bibitem{Arsenault2017}
{Arsenault}, R., {Madec}, P.~Y., {Vernet}, E., {Hackenberg}, W., {La Penna},
  P., {Paufique}, J., {Kuntschner}, H., {Pirard}, J.~F., {Kolb}, J., and
  {Hubin}, N., ``{The Adaptive Optics Facility: Commissioning Progress and
  Results},'' {\em The Messenger}~{\bf 168},  8--14 (June 2017).

\bibitem{Maire2020}
{Maire}, A.-L., {Huby}, E., {Absil}, O., {Zins}, G., {Kasper}, M., {Delacroix},
  C., {Leveratto}, S., {Karlsson}, M., {Ruane}, G., {K{\"a}ufl}, H.-U., {Orban
  de Xivry}, G., {Pathak}, P., {Pettazzi}, L., {Duhoux}, P., {Kolb}, J.,
  {Pantin}, {\'E}., {Riggs}, A.~J.~E., {Siebenmorgen}, R., and {Mawet}, D.,
  ``{Design, pointing control, and on-sky performance of the mid-infrared
  vortex coronagraph for the VLT/NEAR experiment},'' {\em Journal of
  Astronomical Telescopes, Instruments, and Systems}~{\bf 6},  035003 (July
  2020).

\bibitem{Ruane2015}
{Ruane}, G.~J., {Huby}, E., {Absil}, O., {Mawet}, D., {Delacroix}, C.,
  {Carlomagno}, B., and {Swartzlander}, G.~A., ``{Lyot-plane phase masks for
  improved high-contrast imaging with a vortex coronagraph},'' {\em A\&A}~{\bf
  583},  A81 (Nov. 2015).

\bibitem{Ives2014}
{Ives}, D., {Finger}, G., {Jakob}, G., and {Beckmann}, U., ``{AQUARIUS: the
  next generation mid-IR detector for ground-based astronomy, an update.},'' in
  [{\em High Energy, Optical, and Infrared Detectors for Astronomy
  VI}{\nolinebreak\hspace{0.1em}]},  {Holland}, A.~D. and {Beletic}, J., eds.,
  {\em Society of Photo-Optical Instrumentation Engineers (SPIE) Conference
  Series} {\bf 9154},  91541J (July 2014).

\bibitem{Hoffmann2014}
{Hoffmann}, W.~F., {Hinz}, P.~M., {Defr{\`e}re}, D., {Leisenring}, J.~M.,
  {Skemer}, A.~J., {Arbo}, P.~A., {Montoya}, M., and {Mennesson}, B.,
  ``{Operation and performance of the mid-infrared camera, NOMIC, on the Large
  Binocular Telescope},'' in [{\em Ground-based and Airborne Instrumentation
  for Astronomy V}{\nolinebreak\hspace{0.1em}]},  {Ramsay}, S.~K., {McLean},
  I.~S., and {Takami}, H., eds., {\em Society of Photo-Optical Instrumentation
  Engineers (SPIE) Conference Series} {\bf 9147},  91471O (July 2014).

\bibitem{Arrington1998}
{Arrington}, D.~C., {Hubbs}, J.~E., {Gramer}, M.~E., and {Dole}, G.~A.,
  ``{Impact of excess low-frequency noise (ELFN) in Si:As impurity band
  conduction (IBC) focal plane arrays for astronomical applications},'' in
  [{\em Infrared Detectors and Focal Plane Arrays
  V}{\nolinebreak\hspace{0.1em}]},  {Dereniak}, E.~L. and {Sampson}, R.~E.,
  eds., {\em Society of Photo-Optical Instrumentation Engineers (SPIE)
  Conference Series} {\bf 3379},  361--370 (July 1998).

\bibitem{Absil2016}
{Absil}, O., {Mawet}, D., {Karlsson}, M., {Carlomagno}, B., {Christiaens}, V.,
  {Defr{\`e}re}, D., {Delacroix}, C., {Femen{\'\i}a Castella}, B., {Forsberg},
  P., {Girard}, J., {G{\'o}mez Gonz{\'a}lez}, C.~A., {Habraken}, S., {Hinz},
  P.~M., {Huby}, E., {Jolivet}, A., {Matthews}, K., {Milli}, J., {Orban de
  Xivry}, G., {Pantin}, E., {Piron}, P., {Reggiani}, M., {Ruane}, G.~J.,
  {Serabyn}, G., {Surdej}, J., {Tristram}, K. R.~W., {Vargas Catal{\'a}n}, E.,
  {Wertz}, O., and {Wizinowich}, P., ``{Three years of harvest with the vector
  vortex coronagraph in the thermal infrared},'' in [{\em Ground-based and
  Airborne Instrumentation for Astronomy VI}{\nolinebreak\hspace{0.1em}]},
  {Evans}, C.~J., {Simard}, L., and {Takami}, H., eds., {\em Society of
  Photo-Optical Instrumentation Engineers (SPIE) Conference Series} {\bf 9908},
   99080Q (Aug. 2016).

\bibitem{Snik2012}
{Snik}, F., {Otten}, G., {Kenworthy}, M., {Miskiewicz}, M., {Escuti}, M.,
  {Packham}, C., and {Codona}, J., ``{The vector-APP: a broadband apodizing
  phase plate that yields complementary PSFs},'' in [{\em Modern Technologies
  in Space- and Ground-based Telescopes and Instrumentation
  II}{\nolinebreak\hspace{0.1em}]},  {Navarro}, R., {Cunningham}, C.~R., and
  {Prieto}, E., eds., {\em Society of Photo-Optical Instrumentation Engineers
  (SPIE) Conference Series} {\bf 8450},  84500M (Sept. 2012).

\bibitem{Doelman2021}
{Doelman}, D.~S., {Snik}, F., {Por}, E.~H., {Bos}, S.~P., {Otten}, G.~P.~P.~L.,
  {Kenworthy}, M., {Haffert}, S.~Y., {Wilby}, M., {Bohn}, A.~J., {Sutlieff},
  B.~J., {Miller}, K., {Ouellet}, M., {de Boer}, J., {Keller}, C.~U., {Escuti},
  M.~J., {Shi}, S., {Warriner}, N.~Z., {Hornburg}, K., {Birkby}, J.~L.,
  {Males}, J., {Morzinski}, K.~M., {Close}, L.~M., {Codona}, J., {Long}, J.,
  {Schatz}, L., {Lumbres}, J., {Rodack}, A., {Van Gorkom}, K., {Hedglen}, A.,
  {Guyon}, O., {Lozi}, J., {Groff}, T., {Chilcote}, J., {Jovanovic}, N.,
  {Thibault}, S., {de Jonge}, C., {Allain}, G., {Vall{\'e}e}, C., {Patel}, D.,
  {C{\^o}t{\'e}}, O., {Marois}, C., {Hinz}, P., {Stone}, J., {Skemer}, A.,
  {Briesemeister}, Z., {Boehle}, A., {Glauser}, A.~M., {Taylor}, W., {Baudoz},
  P., {Huby}, E., {Absil}, O., {Carlomagno}, B., and {Delacroix}, C.,
  ``{Vector-apodizing phase plate coronagraph: design, current performance, and
  future development [Invited]},'' {\em Applied Optics}~{\bf 60},  D52 (July
  2021).

\bibitem{Huby2016}
Huby, E., Absil, O., Mawet, D., Baudoz, P., Femenìa~Castellà, B., Bottom, M.,
  Ngo, H., and Serabyn, E., ``The qacits pointing sensor: from theory to on-sky
  operation on keck/nirc2,'' {\em Adaptive Optics Systems V}  (Jul 2016).

\bibitem{Huby2017}
Huby, E., Bottom, M., Femenia, B., Ngo, H., Mawet, D., Serabyn, E., and Absil,
  O., ``On-sky performance of the qacits pointing control technique with the
  keck/nirc2 vortex coronagraph,'' {\em A\&A}~{\bf 600},  A46 (Mar 2017).

\bibitem{Marois2006}
{Marois}, C., {Lafreni{\`e}re}, D., {Doyon}, R., {Macintosh}, B., and {Nadeau},
  D., ``{Angular Differential Imaging: A Powerful High-Contrast Imaging
  Technique},'' {\em ApJ}~{\bf 641},  556--564 (Apr. 2006).

\bibitem{Turchi2019}
{Turchi}, A., {Masciadri}, E., {Kerber}, F., and {Martelloni}, G.,
  ``{Forecasting water vapour above the sites of ESO's Very Large Telescope
  (VLT) and the Large Binocular Telescope (LBT)},'' {\em MNRAS}~{\bf 482},
  206--218 (Jan. 2019).

\bibitem{Hinz2016}
{Hinz}, P.~M., {Defr{\`e}re}, D., {Skemer}, A., {Bailey}, V., {Stone}, J.,
  {Spalding}, E., {Vaz}, A., {Pinna}, E., {Puglisi}, A., {Esposito}, S.,
  {Montoya}, M., {Downey}, E., {Leisenring}, J., {Durney}, O., {Hoffmann}, W.,
  {Hill}, J., {Millan-Gabet}, R., {Mennesson}, B., {Danchi}, W., {Morzinski},
  K., {Grenz}, P., {Skrutskie}, M., and {Ertel}, S., ``{Overview of LBTI: a
  multipurpose facility for high spatial resolution observations},'' in [{\em
  Optical and Infrared Interferometry and Imaging
  V}{\nolinebreak\hspace{0.1em}]},  {Malbet}, F., {Creech-Eakman}, M.~J., and
  {Tuthill}, P.~G., eds., {\em Society of Photo-Optical Instrumentation
  Engineers (SPIE) Conference Series} {\bf 9907},  990704 (Aug. 2016).

\bibitem{Ertel2020}
{Ertel}, S., {Hinz}, P.~M., {Stone}, J.~M., {Vaz}, A., {Montoya}, O.~M.,
  {West}, G.~S., {Durney}, O., {Grenz}, P., {Spalding}, E.~A., {Leisenring},
  J., {Wagner}, K., {Anugu}, N., {Power}, J., {Maier}, E.~R., {Defr{\`e}re},
  D., {Hoffmann}, W., {Perera}, S., {Brown}, S., {Skemer}, A.~J., {Mennesson},
  B., {Kennedy}, G., {Downey}, E., {Hill}, J., {Pinna}, E., {Puglisi}, A., and
  {Rossi}, F., ``{Overview and prospects of the LBTI beyond the completed HOSTS
  survey},'' in [{\em Society of Photo-Optical Instrumentation Engineers (SPIE)
  Conference Series}{\nolinebreak\hspace{0.1em}]},  {\em Society of
  Photo-Optical Instrumentation Engineers (SPIE) Conference Series} {\bf
  11446},  1144607 (Dec. 2020).

\bibitem{Stone2018}
{Stone}, J.~M., {Skemer}, A.~J., {Hinz}, P.~M., {Bonavita}, M., {Kratter},
  K.~M., {Maire}, A.-L., {Defrere}, D., {Bailey}, V.~P., {Spalding}, E.,
  {Leisenring}, J.~M., {Desidera}, S., {Bonnefoy}, M., {Biller}, B.,
  {Woodward}, C.~E., {Henning}, T., {Skrutskie}, M.~F., {Eisner}, J.~A.,
  {Crepp}, J.~R., {Patience}, J., {Weigelt}, G., {De Rosa}, R.~J., {Schlieder},
  J., {Brandner}, W., {Apai}, D., {Su}, K., {Ertel}, S., {Ward-Duong}, K.,
  {Morzinski}, K.~M., {Schertl}, D., {Hofmann}, K.-H., {Close}, L.~M., {Brems},
  S.~S., {Fortney}, J.~J., {Oza}, A., {Buenzli}, E., and {Bass}, B., ``{The
  LEECH Exoplanet Imaging Survey: Limits on Planet Occurrence Rates under
  Conservative Assumptions},'' {\em AJ}~{\bf 156},  286 (Dec. 2018).

\bibitem{Wagner2019}
{Wagner}, K., {Stone}, J.~M., {Spalding}, E., {Apai}, D., {Dong}, R., {Ertel},
  S., {Leisenring}, J., and {Webster}, R., ``{Thermal Infrared Imaging of MWC
  758 with the Large Binocular Telescope: Planetary-driven Spiral Arms?},''
  {\em ApJ}~{\bf 882},  20 (Sept. 2019).

\bibitem{Wagner2020}
{Wagner}, K., {Stone}, J., {Dong}, R., {Ertel}, S., {Apai}, D., {Doelman}, D.,
  {Bohn}, A., {Najita}, J., {Brittain}, S., {Kenworthy}, M., {Keppler}, M.,
  {Webster}, R., {Mailhot}, E., and {Snik}, F., ``{First Images of the
  Protoplanetary Disk around PDS 201},'' {\em AJ}~{\bf 159},  252 (June 2020).

\bibitem{Ertel2018}
{Ertel}, S., {Defr{\`e}re}, D., {Hinz}, P., {Mennesson}, B., {Kennedy}, G.~M.,
  {Danchi}, W.~C., {Gelino}, C., {Hill}, J.~M., {Hoffmann}, W.~F., {Rieke}, G.,
  {Shannon}, A., {Spalding}, E., {Stone}, J.~M., {Vaz}, A., {Weinberger},
  A.~J., {Willems}, P., {Absil}, O., {Arbo}, P., {Bailey}, V.~P., {Beichman},
  C., {Bryden}, G., {Downey}, E.~C., {Durney}, O., {Esposito}, S., {Gaspar},
  A., {Grenz}, P., {Haniff}, C.~A., {Leisenring}, J.~M., {Marion}, L.,
  {McMahon}, T.~J., {Millan-Gabet}, R., {Montoya}, M., {Morzinski}, K.~M.,
  {Pinna}, E., {Power}, J., {Puglisi}, A., {Roberge}, A., {Serabyn}, E.,
  {Skemer}, A.~J., {Stapelfeldt}, K., {Su}, K.~Y.~L., {Vaitheeswaran}, V., and
  {Wyatt}, M.~C., ``{The HOSTS Survey{\textemdash}Exozodiacal Dust Measurements
  for 30 Stars},'' {\em AJ}~{\bf 155},  194 (May 2018).

\bibitem{Spalding2019}
{Spalding}, E., {Hinz}, P., {Morzinski}, K., {Ertel}, S., {Grenz}, P., {Maier},
  E., {Stone}, J., and {Vaz}, A., ``{Status of commissioning stabilized
  infrared Fizeau interferometry with LBTI},'' in [{\em Society of
  Photo-Optical Instrumentation Engineers (SPIE) Conference
  Series}{\nolinebreak\hspace{0.1em}]},  {\em Society of Photo-Optical
  Instrumentation Engineers (SPIE) Conference Series} {\bf 11117},  111171S
  (Sept. 2019).

\bibitem{Ertel2020b}
{Ertel}, S., {Defr{\`e}re}, D., {Hinz}, P., {Mennesson}, B., {Kennedy}, G.~M.,
  {Danchi}, W.~C., {Gelino}, C., {Hill}, J.~M., {Hoffmann}, W.~F., {Mazoyer},
  J., {Rieke}, G., {Shannon}, A., {Stapelfeldt}, K., {Spalding}, E., {Stone},
  J.~M., {Vaz}, A., {Weinberger}, A.~J., {Willems}, P., {Absil}, O., {Arbo},
  P., {Bailey}, V.~P., {Beichman}, C., {Bryden}, G., {Downey}, E.~C., {Durney},
  O., {Esposito}, S., {Gaspar}, A., {Grenz}, P., {Haniff}, C.~A., {Leisenring},
  J.~M., {Marion}, L., {McMahon}, T.~J., {Millan-Gabet}, R., {Montoya}, M.,
  {Morzinski}, K.~M., {Perera}, S., {Pinna}, E., {Pott}, J.~U., {Power}, J.,
  {Puglisi}, A., {Roberge}, A., {Serabyn}, E., {Skemer}, A.~J., {Su}, K.~Y.~L.,
  {Vaitheeswaran}, V., and {Wyatt}, M.~C., ``{The HOSTS Survey for Exozodiacal
  Dust: Observational Results from the Complete Survey},'' {\em AJ}~{\bf 159},
  177 (Apr. 2020).

\bibitem{Skemer2014}
{Skemer}, A.~J., {Hinz}, P., {Esposito}, S., {Skrutskie}, M.~F., {Defr{\`e}re},
  D., {Bailey}, V., {Leisenring}, J., {Apai}, D., {Biller}, B., {Bonnefoy}, M.,
  {Brandner}, W., {Buenzli}, E., {Close}, L., {Crepp}, J., {De Rosa}, R.~J.,
  {Desidera}, S., {Eisner}, J., {Fortney}, J., {Henning}, T., {Hofmann}, K.-H.,
  {Kopytova}, T., {Maire}, A.-L., {Males}, J.~R., {Millan-Gabet}, R.,
  {Morzinski}, K., {Oza}, A., {Patience}, J., {Rajan}, A., {Rieke}, G.,
  {Schertl}, D., {Schlieder}, J., {Su}, K., {Vaz}, A., {Ward-Duong}, K.,
  {Weigelt}, G., {Woodward}, C.~E., and {Zimmerman}, N., ``{High contrast
  imaging at the LBT: the LEECH exoplanet imaging survey},'' in [{\em Adaptive
  Optics Systems IV}{\nolinebreak\hspace{0.1em}]},  {Marchetti}, E., {Close},
  L.~M., and {Vran}, J.-P., eds., {\em Society of Photo-Optical Instrumentation
  Engineers (SPIE) Conference Series} {\bf 9148},  91480L (July 2014).

\bibitem{Skemer2016}
{Skemer}, A.~J., {Morley}, C.~V., {Zimmerman}, N.~T., {Skrutskie}, M.~F.,
  {Leisenring}, J., {Buenzli}, E., {Bonnefoy}, M., {Bailey}, V., {Hinz}, P.,
  {Defr{\'e}re}, D., {Esposito}, S., {Apai}, D., {Biller}, B., {Brandner}, W.,
  {Close}, L., {Crepp}, J.~R., {De Rosa}, R.~J., {Desidera}, S., {Eisner}, J.,
  {Fortney}, J., {Freedman}, R., {Henning}, T., {Hofmann}, K.-H., {Kopytova},
  T., {Lupu}, R., {Maire}, A.-L., {Males}, J.~R., {Marley}, M., {Morzinski},
  K., {Oza}, A., {Patience}, J., {Rajan}, A., {Rieke}, G., {Schertl}, D.,
  {Schlieder}, J., {Stone}, J., {Su}, K., {Vaz}, A., {Visscher}, C.,
  {Ward-Duong}, K., {Weigelt}, G., and {Woodward}, C.~E., ``{The LEECH
  Exoplanet Imaging Survey: Characterization of the Coldest Directly Imaged
  Exoplanet, GJ 504 b, and Evidence for Superstellar Metallicity},'' {\em
  ApJ}~{\bf 817},  166 (Feb. 2016).

\bibitem{Brandl2021}
{Brandl}, B., {Bettonvil}, F., {van Boekel}, R., {Glauser}, A., {Quanz}, S.,
  {Absil}, O., {Amorim}, A., {Feldt}, M., {Glasse}, A., {G{\"u}del}, M., {Ho},
  P., {Labadie}, L., {Meyer}, M., {Pantin}, E., {van Winckel}, H., and {METIS
  Consortium}, ``{METIS: The Mid-infrared ELT Imager and Spectrograph},'' {\em
  The Messenger}~{\bf 182},  22--26 (Mar. 2021).

\bibitem{Kasper2021}
{Kasper}, M., {Cerpa Urra}, N., {Pathak}, P., {Bonse}, M., {Nousiainen}, J.,
  {Engler}, B., {Heritier}, C.~T., {Kammerer}, J., {Leveratto}, S., {Rajani},
  C., {Bristow}, P., {Le Louarn}, M., {Madec}, P.~Y., {Str{\"o}bele}, S.,
  {Verinaud}, C., {Glauser}, A., {Quanz}, S.~P., {Helin}, T., {Keller}, C.,
  {Snik}, F., {Boccaletti}, A., {Chauvin}, G., {Mouillet}, D., {Kulcs{\'a}r},
  C., and {Raynaud}, H.~F., ``{PCS {\textemdash} A Roadmap for Exoearth Imaging
  with the ELT},'' {\em The Messenger}~{\bf 182},  38--43 (Mar. 2021).

\bibitem{Blain2018}
{Blain}, C., {Marois}, C., {Bradley}, C., {Chun}, M., {Doyon}, R., {Erickson},
  D., {Hayward}, T., {Lamb}, M., {Lardi{\`e}re}, O., {Marchis}, F., {Melis},
  C., {Meyer}, M., {Packham}, C., {Skemer}, A., and {Thibault}, S., ``{TIKI: a
  10-micron Earth-like planet finder for the Gemini South telescope},'' in
  [{\em Ground-based and Airborne Instrumentation for Astronomy
  VII}{\nolinebreak\hspace{0.1em}]},  {Evans}, C.~J., {Simard}, L., and
  {Takami}, H., eds., {\em Society of Photo-Optical Instrumentation Engineers
  (SPIE) Conference Series} {\bf 10702},  107024A (Aug. 2018).

\bibitem{Morzinski2020}
{Morzinski}, K.~M., {Montoya}, M., {Fellows}, C., {Durney}, O., {Ford}, J.,
  {West}, G., {Gardner}, A., {Vaz}, A., {Anugu}, N., {Mailhot}, E., {Carlson},
  J., {Harrison}, L., {Gacon}, F., {Downey}, E., {Hinz}, P., {Jones}, T.,
  {Patience}, J., {Sivanandam}, S., {Chen}, S., {Lamb}, M., {Butko}, A., {Liu},
  S., {Hardy}, T., and {Jannuzi}, B., ``{Development and status of MAPS, the
  MMT AO exoPlanet characterization system},'' in [{\em Society of
  Photo-Optical Instrumentation Engineers (SPIE) Conference
  Series}{\nolinebreak\hspace{0.1em}]},  {\em Society of Photo-Optical
  Instrumentation Engineers (SPIE) Conference Series} {\bf 11448},  114481L
  (Dec. 2020).

\bibitem{Roman}
{Carri{\'o}n-Gonz{\'a}lez}, {\'O}., {Garc{\'\i}a Mu{\~n}oz}, A., {Cabrera}, J.,
  {Csizmadia}, S., {Santos}, N.~C., and {Rauer}, H., ``{Catalogue of exoplanets
  accessible in reflected starlight to the Nancy Grace Roman Space Telescope. A
  population study and prospects for phase-curve measurements},'' {\em arXiv
  e-prints} ,  arXiv:2104.04296 (Apr. 2021).

\bibitem{Mawet2019}
{Mawet}, D., {Hirsch}, L., {Lee}, E.~J., {Ruffio}, J.-B., {Bottom}, M.,
  {Fulton}, B.~J., {Absil}, O., {Beichman}, C., {Bowler}, B., {Bryan}, M.,
  {Choquet}, E., {Ciardi}, D., {Christiaens}, V., {Defr{\`e}re}, D., {Gomez
  Gonzalez}, C.~A., {Howard}, A.~W., {Huby}, E., {Isaacson}, H., {Jensen-Clem},
  R., {Kosiarek}, M., {Marcy}, G., {Meshkat}, T., {Petigura}, E., {Reggiani},
  M., {Ruane}, G., {Serabyn}, E., {Sinukoff}, E., {Wang}, J., {Weiss}, L., and
  {Ygouf}, M., ``{Deep Exploration of $\epsilon$ Eridani with Keck Ms-band
  Vortex Coronagraphy and Radial Velocities: Mass and Orbital Parameters of the
  Giant Exoplanet},'' {\em AJ}~{\bf 157},  33 (Jan. 2019).

\bibitem{Pathak2021}
{Pathak}, P., {Petit dit de la Roche}, D.~J.~M., {Kasper}, M., {Sterzik}, M.,
  {Absil}, O., {Boehle}, A., {Feng}, F., {Ivanov}, V.~D., {Janson}, M.,
  {Jones}, H.~R.~A., {Kaufer}, A., {K{\"a}ufl}, H.~U., {Maire}, A.~L., {Meyer},
  M., {Pantin}, E., {Siebenmorgen}, R., {van den Ancker}, M.~E., and
  {Viswanath}, G., ``{High contrast imaging at 10 microns, a search for
  exoplanets around: Eps Indi A, Eps Eri, Tau Ceti, Sirius A and Sirius B},''
  {\em arXiv e-prints} ,  arXiv:2104.13032 (Apr. 2021).

\bibitem{Viswanath2021}
{Viswanath}, G., {Janson}, M., {Dahlqvist}, C.-H., {Petit dit de la Roche}, D.,
  {Samland}, M., {Girard}, J., {Pathak}, P., {Kasper}, M., {Feng}, F., {Meyer},
  M., {Boehle}, A., {Quanz}, S.~P., {Jones}, H. R.~A., {Absil}, O., {Brandner},
  W., {Maire}, A.-L., {Siebenmorgen}, R., {Sterzik}, M., and {Pantin}, E.,
  ``{Constraints on the nearby exoplanet $\epsilon$ Ind Ab from deep
  near/mid-infrared imaging limits},'' {\em arXiv e-prints} ,  arXiv:2105.09773
  (May 2021).

\bibitem{Wagner2019b}
{Wagner}, K., {Apai}, D., and {Kratter}, K.~M., ``{On the Mass Function,
  Multiplicity, and Origins of Wide-orbit Giant Planets},'' {\em ApJ}~{\bf
  877},  46 (May 2019).

\bibitem{Nielsen2019}
{Nielsen}, E.~L., {De Rosa}, R.~J., {Macintosh}, B., {Wang}, J.~J., {Ruffio},
  J.-B., {Chiang}, E., {Marley}, M.~S., {Saumon}, D., {Savransky}, D.,
  {Ammons}, S.~M., {Bailey}, V.~P., {Barman}, T., {Blain}, C., {Bulger}, J.,
  {Burrows}, A., {Chilcote}, J., {Cotten}, T., {Czekala}, I., {Doyon}, R.,
  {Duch{\^e}ne}, G., {Esposito}, T.~M., {Fabrycky}, D., {Fitzgerald}, M.~P.,
  {Follette}, K.~B., {Fortney}, J.~J., {Gerard}, B.~L., {Goodsell}, S.~J.,
  {Graham}, J.~R., {Greenbaum}, A.~Z., {Hibon}, P., {Hinkley}, S., {Hirsch},
  L.~A., {Hom}, J., {Hung}, L.-W., {Dawson}, R.~I., {Ingraham}, P., {Kalas},
  P., {Konopacky}, Q., {Larkin}, J.~E., {Lee}, E.~J., {Lin}, J.~W., {Maire},
  J., {Marchis}, F., {Marois}, C., {Metchev}, S., {Millar-Blanchaer}, M.~A.,
  {Morzinski}, K.~M., {Oppenheimer}, R., {Palmer}, D., {Patience}, J.,
  {Perrin}, M., {Poyneer}, L., {Pueyo}, L., {Rafikov}, R.~R., {Rajan}, A.,
  {Rameau}, J., {Rantakyr{\"o}}, F.~T., {Ren}, B., {Schneider}, A.~C.,
  {Sivaramakrishnan}, A., {Song}, I., {Soummer}, R., {Tallis}, M., {Thomas},
  S., {Ward-Duong}, K., and {Wolff}, S., ``{The Gemini Planet Imager Exoplanet
  Survey: Giant Planet and Brown Dwarf Demographics from 10 to 100 au},'' {\em
  AJ}~{\bf 158},  13 (July 2019).

\bibitem{Vigan2020}
{Vigan}, A., {Fontanive}, C., {Meyer}, M., {Biller}, B., {Bonavita}, M.,
  {Feldt}, M., {Desidera}, S., {Marleau}, G.~D., {Emsenhuber}, A., {Galicher},
  R., {Rice}, K., {Forgan}, D., {Mordasini}, C., {Gratton}, R., {Le Coroller},
  H., {Maire}, A.~L., {Cantalloube}, F., {Chauvin}, G., {Cheetham}, A.,
  {Hagelberg}, J., {Lagrange}, A.~M., {Langlois}, M., {Bonnefoy}, M., {Beuzit},
  J.~L., {Boccaletti}, A., {D'Orazi}, V., {Delorme}, P., {Dominik}, C.,
  {Henning}, T., {Janson}, M., {Lagadec}, E., {Lazzoni}, C., {Ligi}, R.,
  {Menard}, F., {Mesa}, D., {Messina}, S., {Moutou}, C., {M{\"u}ller}, A.,
  {Perrot}, C., {Samland}, M., {Schmid}, H.~M., {Schmidt}, T., {Sissa}, E.,
  {Turatto}, M., {Udry}, S., {Zurlo}, A., {Abe}, L., {Antichi}, J.,
  {Asensio-Torres}, R., {Baruffolo}, A., {Baudoz}, P., {Baudrand}, J.,
  {Bazzon}, A., {Blanchard}, P., {Bohn}, A.~J., {Brown Sevilla}, S.,
  {Carbillet}, M., {Carle}, M., {Cascone}, E., {Charton}, J., {Claudi}, R.,
  {Costille}, A., {De Caprio}, V., {Delboulb{\'e}}, A., {Dohlen}, K., {Engler},
  N., {Fantinel}, D., {Feautrier}, P., {Fusco}, T., {Gigan}, P., {Girard},
  J.~H., {Giro}, E., {Gisler}, D., {Gluck}, L., {Gry}, C., {Hubin}, N.,
  {Hugot}, E., {Jaquet}, M., {Kasper}, M., {Le Mignant}, D., {Llored}, M.,
  {Madec}, F., {Magnard}, Y., {Martinez}, P., {Maurel}, D.,
  {M{\"o}ller-Nilsson}, O., {Mouillet}, D., {Moulin}, T., {Orign{\'e}}, A.,
  {Pavlov}, A., {Perret}, D., {Petit}, C., {Pragt}, J., {Puget}, P., {Rabou},
  P., {Ramos}, J., {Rickman}, E.~L., {Rigal}, F., {Rochat}, S., {Roelfsema},
  R., {Rousset}, G., {Roux}, A., {Salasnich}, B., {Sauvage}, J.~F., {Sevin},
  A., {Soenke}, C., {Stadler}, E., {Suarez}, M., {Wahhaj}, Z., {Weber}, L., and
  {Wildi}, F., ``{The SPHERE infrared survey for exoplanets (SHINE). III. The
  demographics of young giant exoplanets below 300 au with SPHERE},'' {\em
  arXiv e-prints} ,  arXiv:2007.06573 (July 2020).

\bibitem{Fernandes2019}
{Fernandes}, R.~B., {Mulders}, G.~D., {Pascucci}, I., {Mordasini}, C., and
  {Emsenhuber}, A., ``{Hints for a Turnover at the Snow Line in the Giant
  Planet Occurrence Rate},'' {\em ApJ}~{\bf 874},  81 (Mar. 2019).

\bibitem{Fulton2021}
{Fulton}, B.~J., {Rosenthal}, L.~J., {Hirsch}, L.~A., {Isaacson}, H., {Howard},
  A.~W., {Dedrick}, C.~M., {Sherstyuk}, I.~A., {Blunt}, S.~C., {Petigura},
  E.~A., {Knutson}, H.~A., {Behmard}, A., {Chontos}, A., {Crepp}, J.~R.,
  {Crossfield}, I. J.~M., {Dalba}, P.~A., {Fischer}, D.~A., {Henry}, G.~W.,
  {Kane}, S.~R., {Kosiarek}, M., {Marcy}, G.~W., {Rubenzahl}, R.~A., {Weiss},
  L.~M., and {Wright}, J.~T., ``{The California Legacy Survey II. Occurrence of
  Giant Planets Beyond the Ice line},'' {\em arXiv e-prints} ,
  arXiv:2105.11584 (May 2021).

\bibitem{Males2015}
{Males}, J.~R., {Belikov}, R., and {Bendek}, E., ``{Orbital Differential
  Imaging: a new high-contrast post-processing technique for direct imaging of
  exoplanets},'' in [{\em Techniques and Instrumentation for Detection of
  Exoplanets VII}{\nolinebreak\hspace{0.1em}]},  {Shaklan}, S., ed., {\em
  Society of Photo-Optical Instrumentation Engineers (SPIE) Conference Series}
  {\bf 9605},  960518 (Sept. 2015).

\bibitem{LeCoroller2020}
{Le Coroller}, H., {Nowak}, M., {Delorme}, P., {Chauvin}, G., {Gratton}, R.,
  {Devinat}, M., {Bec-Canet}, J., {Schneeberger}, A., {Estevez}, D., {Arnold},
  L., {Beust}, H., {Bonnefoy}, M., {Boccaletti}, A., {Desgrange}, C.,
  {Desidera}, S., {Galicher}, R., {Lagrange}, A.~M., {Langlois}, M., {Maire},
  A.~L., {Menard}, F., {Vernazza}, P., {Vigan}, A., {Zurlo}, A., {Fenouillet},
  T., {Lambert}, J.~C., {Bonavita}, M., {Cheetham}, A., {D'orazi}, V., {Feldt},
  M., {Janson}, M., {Ligi}, R., {Mesa}, D., {Meyer}, M., {Samland}, M.,
  {Sissa}, E., {Beuzit}, J.~L., {Dohlen}, K., {Fusco}, T., {Le Mignant}, D.,
  {Mouillet}, D., {Ramos}, J., {Rochat}, S., and {Sauvage}, J.~F.,
  ``{K-Stacker: an algorithm to hack the orbital parameters of planets hidden
  in high-contrast imaging. First applications to VLT/SPHERE multi-epoch
  observations},'' {\em A\&A}~{\bf 639},  A113 (July 2020).

\bibitem{Fitzgerald2019}
{Fitzgerald}, M., {Bailey}, V., {Baranec}, C., {Batalha}, N., {Benneke}, B.,
  {Beichman}, C., {Brandt}, T., {Chilcote}, J., {Chun}, M., {Crossfield}, I.,
  {Currie}, T., {Davis}, K., {Dekany}, R., {Delorme}, J.-R., {Dong}, R.,
  {Doyon}, R., {Dressing}, C., {Echeverri}, D., {Fortney}, J., {Frazin}, R.~A.,
  {Guyon}, O., {Hashimoto}, J., {Hillenbrand}, L., {Hinz}, P., {Howard}, A.,
  {Jensen-Clem}, R., {Jovanovic}, N., {Kawahara}, H., {Knutson}, H.,
  {Konopacky}, Q., {Kotani}, T., {Lafreni{\`e}re}, D., {Liu}, M., {Lozi}, J.,
  {Lu}, J.~R., {Males}, J., {Marley}, M., {Marois}, C., {Mawet}, D., {Mazin},
  B., {Millar-Blanchaer}, M., {Mondal}, S., {Murakami}, N., {Murray-Clay}, R.,
  {Narita}, N., {Pezzato}, J., {Pyo}, T.-S., {Roberts}, L., {Ruane}, G.,
  {Sallum}, S., {Serabyn}, G., {Shields}, A., {Simard}, L., {Skemer}, A.,
  {Stelter}, R.~D., {Tamura}, M., {Troy}, M., {Vasisht}, G., {Wallace}, J.~K.,
  {Wang}, J., {Wang}, J., and {Wright}, S.~A., ``{The Planetary Systems Imager
  for TMT},'' in [{\em Bulletin of the American Astronomical
  Society}{\nolinebreak\hspace{0.1em}]},   {\bf 51},  251 (Sept. 2019).

\end{thebibliography}
\bibliographystyle{spiebib} 

\end{document}